\documentclass{ws-procs975x65}

\newcommand{\bfb}[1]{\mbox{\boldmath $ #1 $}}

\begin{document}

\title{Nonlinear Schr\"{o}dinger-Pauli Equations}

\author{Wei Khim Ng$^*$ and Rajesh R. Parwani$^\dag$}

\address{Department of Physics, National University of Singapore,\\
2 Science Drive 3 Singapore 117542\\
$^*$phynwk@nus.edu.sg,
$^\dag$parwani@nus.edu.sg}

\begin{abstract}
We obtain novel nonlinear Schr\"{o}dinger-Pauli equations through a formal non-relativistic limit of appropriately constructed nonlinear Dirac equations. This procedure automatically provides a physical regularisation of potential singularities brought forward by the nonlinear terms and suggests how to regularise previous equations studied in the literature. The enhancement of contributions coming from the regularised singularities suggests that the obtained equations might be useful for future precision tests of quantum nonlinearity.\end{abstract}


\bodymatter

\section{Introduction}\label{intro}
Several nonlinear extensions of Schr\"{o}dinger's equation have been constructed to probe the accuracy of quantum
linearity \cite{Bial,kibble,weinberg}. For example, Weinberg proposed a class of equations which were then used
in several experimental tests, see \cite{Expts,chupp,walsworth,majumder} and references therein. The results
indicated that any potential non-linearity in those systems had to be smaller than some bounds.

Ignoring external fields, the nonlinear Schr\"{o}dinger equations may be written in the form
\begin{equation}
 i\hbar\frac{\partial}{\partial t}\psi = -\frac{\hbar^2}{2m}{\nabla}^2 \psi  + f_{NR}(\psi) \psi
    \end{equation}
where the nonlinearity $f_{NR}$ depends in general on the wavefunction, its conjugate and their derivatives.
$f_{NR}$ may be written as a ratio of two terms, $N(\psi)/D(\psi)$, with equal factors of $\psi$ in the
numerator and denominator to keep the scale invariance $\psi \to \lambda \psi$, $\lambda$ a constant,  of the linear
Schr\"{o}dinger equation. This scale invariance ensures the wavefunction can be freely normalised. The denominator is typically a monomial in $\psi^{\star} \psi$ so that the nonlinear term
may be made separable  for independent systems. As the nonlinearity must be weak on phenomenological grounds,
the solutions of the linear equation must be very close to some solutions of the modified equation. However, any
solutions of the linear equation that have nodes would make $D(\psi)$ vanish at some points. Thus, the
nonlinearity would generally be singular and ill-defined at those points.

Weinberg\cite{weinberg}, discusses classes of nonlinear Schr\"{o}dinger-Pauli equations where the nonlinearity
turns out to be finite at the nodes because the numerator vanishes faster than the denominator there. However
this will not happen for general classes of nonlinearities where $N(\psi)$ has derivatives, such as the
equations studied in Ref\cite{dgn}.

It was suggested in Ref. \cite{RP2} that quantum nonlinearity might be linked to the breaking of space-time symmetry. This idea was supported by a study in Ref. \cite{NP3} in the relativistic regime: That is, a deviation from quantum linearity is associated with a violation of Lorentz symmetry. Thus, in this paper, we discuss how to construct novel classes of nonlinear Schr\"{o}dinger-Pauli equations, which have the above-mentioned scale invariance, starting from nonlinear Dirac equations. However, we will keep our option open by considering both Lorentz invariant and Lorentz violating nonlinear Dirac equations \cite{NP1}. 

As we shall see, our procedure of obtaining the Scrodinger-Pauli equations by taking the limit of relativistic equations has the advantage of indicating a natural and physical regularisation of the singularities. We remark that we focus on genuine nonlinear Dirac equations that cannot be linearised by performing a nonlinear gauge transformation \cite{NP1}.

In the next section, we discuss, in general terms, the formal non-relativistic limit of a subset of nonlinear
Dirac equations constructed in Ref\cite{NP1}. For conciseness, in this paper we only consider the case where $F=fI$
in (\ref{nld}), $I$ being the identity matrix in spinor space. Explicit examples of the lowest order
nonlinearities, corresponding to one factor of $\psi^{\star} \psi$ in $D(\psi)$ are exhibited in Section
\ref{eg}, other cases being similarly handled. The singularity resolution is discussed in Section \ref{apparent
sing} and we end with a discussion in Section \ref{disc}.

We note in passing that nonlinear Schr\"{o}dinger equations of other types have been constructed from Levy-Leblond's
``non-relativistic Dirac equation" which is itself the non-relativistic limit of the usual Dirac equation
\cite{duval1,duval2}.

\section{Non-Relativistic Limit}\label{nr}
We start from nonlinear Dirac equations of the form
\begin{equation}\label{nld}
        \left(i\hbar\gamma^\mu\partial_\mu-mc + \epsilon F \right) \psi=0 \, ,
\end{equation}
where $F=F(\psi,\bar{\psi})=fI$ and where we have made the small parameter $\epsilon$ explicit. We demand that
$F$ has certain properties so that desirable characteristics of the linear Dirac equation, such as locality,
conservation of probability, separability and invariance under $\psi \to \lambda \psi$, are retained (we are
adopting the standard kinematical structure of quantum mechanics, in particular the standard inner product).
 The other symbols in (\ref{nld}), such as those for the gamma matrices, have their usual meanings; our conventions are similar to those in the textbook \cite{IZ} and in Ref \cite{NP1}.

In Hamiltonian form the equation is
\begin{eqnarray}\label{nr0.3} 
i\hbar\frac{\partial}{\partial t}\psi&=&\left(i\hbar c\bfb{\alpha}\cdot\bfb{\nabla}+\beta mc^2- \epsilon c \beta  f\right) \psi
\end{eqnarray}
where $\alpha^i=\gamma^0\gamma^i$ and $\beta=\gamma^0$. It maybe be decomposed  into two equations by
introducing upper and lower components of the wavefunction,
     \begin{equation}\label{nr0.4}
        \psi= \left(\begin{array}{c}\varphi\\\chi\end{array}\right)e^{-imc^2t/\hbar}
    \end{equation}
 where the rest energy has been extracted as it is the largest component in the non-relativistic limit. We adopt the standard textbook procedure in obtaining the leading non-relativistic limit, but for clarity we repeat some steps below. In order to make the algebra manageable, we simply take $1/c$ to be the same order of magnitude as the nonlinearity scale $\epsilon$ and keep only the leading nonlinear term in the standard non-relativistic expansion. Thus we can isolate the leading order
nonlinear contribution. However, in realistic applications, $\epsilon$ will be much smaller than $1/c$: This will introduce higher order, $1/c$, terms which \emph{do not affect the leading order nonlinear contribution}.

Substituting (\ref{nr0.4}) into (\ref{nr0.3}) we get
    \begin{equation}\label{nr0.7}
        i\hbar\frac{\partial}{\partial
        t}\left(\begin{array}{c}\varphi\\\chi\end{array}\right)=i\hbar
        c\left(\begin{array}{c}\bfb{\sigma}\cdot\bfb{\nabla}\chi\\\bfb{\sigma}\cdot\bfb{\nabla}\varphi\end{array}\right)+mc^2\left(\begin{array}{c}0\\-2\chi\end{array}\right)- \epsilon  cf \left(\begin{array}{c}\varphi\\-\chi\end{array}\right) \, .
    \end{equation}
(As in the usual textbook procedure, the {\it ansatz} (\ref{nr0.4}) removes the mass term for the upper
component).
    From the lower component of (\ref{nr0.7}) we have,
        \begin{equation}\label{nr0.9}
        \chi=\frac{i\hbar\bfb{\sigma}\cdot\bfb{\nabla}\varphi}{2mc}-\frac{i\hbar}{2mc^2}\frac{\partial\chi}{\partial t}+\frac{\epsilon f \chi}{2mc} \, .
    \end{equation}
    Let
    $\chi_0=\frac{i\hbar\bfb{\sigma}\cdot\bfb{\nabla}\varphi}{2mc}$.
Then expanding  (\ref{nr0.9}) about $\chi_0$, we obtain $\chi=\chi_0+O\left(\frac{\epsilon}{c^2},\frac{1}{c^3}\right)$. That is, $\chi$ is the same as that in the linear
    theory. Substituting (\ref{nr0.9}) into the upper component of (\ref{nr0.7}), we arrive at
    \begin{eqnarray}\label{nr0.11}
                i\hbar\frac{\partial}{\partial t}\varphi&\simeq& -\frac{\hbar^2}{2m}{\nabla}^2 \varphi  -\epsilon cf_{NR} \varphi
    \end{eqnarray}
    where $f_{NR}$ means that the state dependence of $f$ has been simplified using (\ref{nr0.4}, \ref{nr0.9}) and higher order terms dropped. Below we look at some explicit examples.

\section{Examples}\label{eg}

\subsection{Lorentz invariant $f$ with one derivative}\label{eg1}

A Lorentz invariant $f$ with one derivative and which is odd under the parity transformation is
    \begin{equation}\label{nr1.11}
       f_1= \epsilon \frac{\partial_\mu j^\mu_5}{\bar \psi\psi} \, ,
           \end{equation}
    where $j^\mu_5 = \bar \psi\gamma^\mu\gamma_5\psi$ is the usual chiral current. The non-relativistic limit is
   \begin{equation}\label{nr1.12}
        i\hbar\frac{\partial\varphi}{\partial t}\simeq-\frac{\hbar^2\nabla^2\varphi}{2m}-\epsilon c \varphi\frac{\bfb{\nabla}\cdot\left(\varphi^\dag\bfb{\sigma}\varphi\right)}{|\varphi|^2}\, .
    \end{equation}
    The factor $\bfb{\nabla}\cdot\left(\varphi^\dag\bfb{\sigma}\varphi\right)$
    appears often in parity odd equations \cite{NP1}; it couples the spin components of the two-component spinor.

\subsection{Lorentz invariant $f$ with two derivatives}\label{eg2}

    For an example of a Lorentz invariant $f$ with two derivatives consider
    \begin{equation}\label{nr2.1}
       f_2=\frac{\epsilon\left(\partial_\mu\partial^\mu\bar \psi\psi\right)+ \delta\left(\partial_\mu\bar \psi\right)\left(\partial^\mu\psi\right)}{\bar \psi\psi} \, ,
    \end{equation}
    where $\epsilon$ and $\delta$ are two independent small parameters taken to be of order $1/c$ below. The non-relativistic equation is
       \begin{eqnarray}\label{nr2.9}
        i\hbar\frac{\partial\varphi}{\partial t}&\simeq&-\frac{\hbar^2\nabla^2\varphi}{2m}\left(1+\frac{mc\delta}{2\hbar^2}\right)+\frac{\varphi}{|\varphi|^2}\left\{-\frac{\delta imc}{\hbar}\left[\varphi^\dag\frac{\partial\varphi}{\partial t}-\left(\frac{\partial\varphi^\dag}{\partial t}\right)\varphi\right]\right.\nonumber\\
                    &&+\left.\epsilon c\left(\nabla^2\varphi^\dag\varphi\right)+\delta c\left(\nabla\varphi^\dag\right)\cdot\left(\nabla\varphi\right)\right\}\, .
    \end{eqnarray}

\subsection{Lorentz violating, parity even $f$}\label{eg3}

Lorentz violating non-linear Dirac equations are of some interest \cite{NP1,RP2,NP2,Carr,nwk}. An example of such an
$f$ with no derivatives and even under parity is
        \begin{equation}\label{nr3.1}
         f_3=A_\mu\frac{\bar \psi\gamma^\mu\psi}{\bar \psi\psi}
    \end{equation}
    where $A_{\mu}$ is a constant vector background field. The non-relativistic limit is
    \begin{equation}\label{nr3.2}
        i\hbar\frac{\partial\varphi}{\partial t}\simeq-\frac{\hbar^2\nabla^2\varphi}{2m}-cA_0\varphi+\frac{i\hbar\varphi}{2m}\frac{\bfb{A}\cdot\left[\varphi^\dag\nabla\varphi-\left(\nabla\varphi^\dag\right)\varphi\right]}{|\varphi|^2}\, .
    \end{equation}

\subsection{Lorentz violating, parity odd $f$}\label{eg4}

    A Lorentz violating $f$ which is odd under parity is
    \begin{equation}\label{nr3.3}
        f_4=A_\mu\frac{\bar \psi\gamma_5\gamma^\mu\psi}{\bar \psi\psi} \, .
    \end{equation}
    The non-relativistic equation is
   \begin{eqnarray}\label{nr3.4}
        i\hbar\frac{\partial\varphi}{\partial t}&\simeq&-\frac{\hbar^2\nabla^2\varphi}{2m}-\frac{c\varphi^\dag\bfb{A}\cdot\bfb{\sigma}\varphi}{|\varphi|^2}\varphi+\frac{A_0i\hbar\varphi}{2m}\frac{\left[\varphi^\dag\bfb{\sigma}\cdot\nabla\varphi-\left(\nabla\varphi^\dag\right)\cdot\bfb{\sigma}\varphi\right]}{|\varphi|^2} \, .
    \end{eqnarray}

\section{Apparent Singularities}\label{apparent sing}
From the above examples, we see the appearance of the following structures in the non-linear Schr\"{o}dinger-Pauli
equations,
\begin{equation}
X=\frac{\varphi^\dag\bfb{\sigma}\cdot\bfb{\nabla}\varphi}{|\varphi|^2}\,\,\,,\,\,\,Y=\frac{(\bfb{\nabla}\varphi^\dag)\cdot(\bfb{\nabla}\varphi)}{|\varphi|^2}\,\,\,,\,\,\,Z=\frac{\varphi^\dag\bfb\nabla^2\varphi}{|\varphi|^2}
\, .
\end{equation}
Clearly, at the nodes of $\varphi$, these forms are singular. However, we can avoid these singularities in a
natural way. For our nonlinear Dirac equations \cite{NP1}, the nonlinearities have the common structure
$\frac{N(\bar \psi,\psi)}{(\bar\psi \psi)^n}$, the  $n=1$ case being discussed here. In terms of the two
component spinors this is  $\frac{N}{|\varphi|^2-|\chi|^2}$ where the lower (small) component contribution
$|\chi|^2$ is usually dropped in the non-relativistic limit. However, at the nodes of $\varphi$, we must keep
the small component in the denominator. This regulates the above mentioned singularity for the following reason:
From (\ref{nr0.9}), the lower component is proportional to the slope of $\varphi$ (i.e. $\nabla\varphi$), which
is unlikely to vanish simultaneously at the nodes except for special cases. In such extreme cases, one would
need to retain the smaller terms (higher order in $1/c$) in the non-relativistic expansion of the denominator.

For the specific examples illustrated above, the replacement $|\varphi|^2 \to |\varphi|^2-|\chi|^2$ in the
denominator makes $X=Z=0$ at a node of $\varphi$ while $Y$ becomes finite and actually enhanced because of the
small denominator. Note that, at the level of the equation of motion, there is an extra factor of $\varphi$
which multiplies the nonlinearity $f$. It is clear that $X$ and $Z$ contributions in the equation of motion are
not singular even at the nodes but the $Y$ contribution is, unless regularised as discussed above.

So far we have discussed singularities in $f$ and at the level of equations of motion. As for observables, let
us consider shifts in the energy levels given by first-order perturbation theory,
\begin{equation}
\delta E =\int d^3x\,<\varphi|F|\varphi>=\int d^3x|\varphi|^2 f(\varphi)
\end{equation}
where the unperturbed (linear equation) wavefunctions are used. We see that the $X, Y, Z$ structures give finite
shifts. Singularities will appear in $n\geq 2$ classes of nonlinearities discussed in Ref\cite{NP1}, two examples
of which are given by
\begin{eqnarray}
V&=&Y^2=\frac{\left[(\bfb{\nabla}\varphi^\dag)\cdot(\bfb{\nabla}\varphi)\right]\left[(\bfb{\nabla}\varphi^\dag)\cdot(\bfb{\nabla}\varphi)\right]}{|\varphi|^2|\varphi|^2} \, , \\
W&=&YZ=\frac{\left[(\bfb{\nabla}\varphi^\dag)\cdot(\bfb{\nabla}\varphi)\right](\varphi^\dag\nabla^2\varphi)}{|\varphi|^2|\varphi|^2}
\, .
\end{eqnarray}
It is clear that the energy shifts will be singular for such terms unless the regularisation is implemented.

The above discussion has ignored external potentials which must be included in realistic experiments. For
example, in the presence of an external gauge field and for a particular spin component
$\varphi=\left(\begin{array}{c}1\\0\end{array}\right)\varphi_0$, the lower component is modified from its
previous form $\chi_0$ to become
\begin{equation}
\chi_0=\frac{i\hbar}{2mc}\left(\frac{\partial}{\partial z}-\frac{e}{c}A_z\right)\varphi_0 \, .
\end{equation}
Setting $\varphi_0=g(\bfb{x}-\bfb{x_0})$ near a node we have
\begin{equation}
|\chi_0|^2=\frac{\hbar^2}{4m^2c^2}\left[\left(\frac{\partial g}{\partial
z}\right)^2-2\frac{e}{c}A_zg\frac{\partial g}{\partial z}+\frac{e^2}{c^2}A^2_zg^2\right] \, .
\end{equation}
In this case, at the node of $\varphi$, $|\chi_0|^2$ has exactly the same form as the case when the gauge field
is absent.

We can see that the contributions from nonlinear effects are largest (if non-zero) at the nodes. This suggests
that future tests for quantum nonlinearity should focus on systems containing nodes in their wavefunctions.

We remark that the nonlinear equations discussed in Ref.\cite{dgn} have been applied to the hydrogen system \cite{dgn2} but the physical consequences of singularities at the nodes of wavefunctions ($\varphi\rightarrow 0$) was not discussed.

\section{Discussion}\label{disc}
We have illustrated how to obtain novel classes of nonlinear Schr\"{o}dinger-Pauli equations starting from the
nonlinear Dirac equations constructed in Ref\cite{NP1}, the latter equations  themselves being more general than
previous constructions \cite{doebner,NLDS1,NLDS2}. For example, we have cases where the time-derivatives appear
in the nonlinearity, and cases where the two components of the spinor are coupled through parity violation. We
remark that probability is conserved for all of our non-relativistic equations. Also, the equations that are
descended from Lorentz covariant equations are Galilean invariant.

An interesting point to note is that certain Lorentz-violating nonlinear Dirac equations have non-relativistic
limits that are Galilean invariant. For example, for $f_3$, if the  background field has only a time component,
the leading non-relativistic limit actually becomes linear and invariant under Galilean transformations. For
$f_4$, choosing a space-like background field will cause the non-relativistic equation to be still nonlinear but
invariant under Galilean transformations.

We had taken the nonlinearity parameter $\epsilon$ to be the same order of magnitude as $1/c$ for ease of power
counting, as our main aim was to isolate the leading nonlinear structure in the formal non-relativistic limit.
We saw that potential singularities in the Schr\"{o}dinger-Pauli equations are regularised by keeping the subleading
lower components of the four component Dirac spinor in the denominators of the nonlinear terms. Thus,
physically, it is the relativistic corrections that regulate the singularities. Precisely at a node, if the
numerator is is nonzero, the nonlinearity is actually enhanced  by the small denominator.

The situation here is qualitatively similar to a previous study \cite{PT} of an information-theoretic
motivated nonlinear Schr\"{o}dinger equation \cite{RP2}, where the contribution to energy shifts from states with
nodes was enhanced relative to states which had no nodes. Note also that in replacing the potentially singular
denominator $|\varphi|^2$ by $|\varphi|^2-|\chi|^2$ as in Section \ref{apparent sing}, one has introduced an
infinite number of derivatives, through a formal expansion of the denominator, into the nonlinear terms even
though we had started with a finite number of derivatives. This again is qualitatively similar to the situation
with the information-theoretic nonlinearity\cite{RP2}.

In actual applications, such as tests of quantum linearity, one would have to set $\epsilon$ much smaller than
$1/c$ in the constructed nonlinear Schr\"{o}dinger-Pauli equations even though they were formally derived from the
nonlinear Dirac equations assuming $\epsilon \sim 1/c$.

The main suggestion from this study is that future precision low-energy experiments, probing deviations from
quantum linearity, should focus on systems which have nodes in their limiting linear wavefunctions. It is there
that the nonlinearity, if nonzero, will be enhanced.

\end{document}